\documentclass[a4paper,11pt]{article}
\usepackage{pos}
\usepackage{lineno}
\usepackage[normalem]{ulem}

\title{Searching for Lorentz invariance violation with artificial neural networks}

\author*[a]{Tomislav Terzić}
\author[a]{Karlo Mrakovčić}

\affiliation[a]{University of Rijeka, Faculty of Physics\\
  Radmile Matejčić 2, Rijeka, Croatia}

\emailAdd{tomislav.terzic@gmail.com}
\emailAdd{karlo.mrakovcic@phy.uniri.hr}

\abstract{Lorentz invariance violation (LIV) can have multiple consequences on very-high energy gamma rays' emission, propagation, and detection, such as energy-dependent photon group velocity, photon instability, vacuum birefringence, and modified electromagnetic interaction. Depending on the underlying theoretical model, several of these effects can coexist. Nevertheless, in experimental tests of LIV, each effect is tested separately and independently. Here, we are performing a search for traces of several coexisting effects in a single analysis. We present our analysis method based on artificial neural networks and put our very first results in the context of experimental searches for LIV.}

\FullConference{
}

\begin{document}
\maketitle

\section{Introduction}
\label{sec:intro}

The search for the theory of quantum gravity (QG) is hampered by the lack of experimental evidence. The Planck energy scale ($E_\mathrm{Pl} \approx 1.2 \times 10^{19}$\,GeV), as the expected domain of QG is far beyond the reach of accelerator experiments. At the end of $20^{\text{th}}$ century, an idea appeared that highly energetic particles from astrophysical sources might be able to probe spacetime fluctuations, expected to occur at the Planck scale distances~\cite{AmelinoCamelia:1997gz}. Gamma rays are particularly good probes because they are detected with relatively high statistics, reach energies of up to a PeV scale~\cite{LHAASO:2021gok}, and propagate on straight lines, which facilitates source identification. 

The interaction of gamma rays with spacetime is expected to manifest itself as energy-dependent photon group velocity. This is a possible consequence of Lorentz invariance violation (LIV). Other possible effects of LIV include modified reaction thresholds, as well as modified interaction dynamics, vacuum birefringence, etc. A comprehensive review of different QG models and tests of LIV using cosmic messengers (cosmic rays, gamma rays, gravitational waves, neutrinos) can be found in~\cite{Addazi:2021xuf, PerezdelosHeros:2022izj}, while~\cite{Terzic:2021rlx} offers a critical discussion and comparison of tests and analysis methods performed using gamma rays. 

Depending on the underlying theoretical model, several LIV effects can manifest simultaneously. However, testing multiple LIV effects simultaneously remains a major challenge. Most of the existing analyses focus on individual phenomena---such as energy-dependent time delays or modified reaction---in isolation. Here, we explore a joint approach using two distinct machine learning architectures: a dense feedforward artificial neural network (ANN) based on parameterized fits of observed data, and a  sequence-to-regression transformer trained on full-photon time-energy sequences.

\section{Theoretical framework}
\label{sec:theory}

LIV is typically parametrized through a modified photon dispersion relation:
\begin{equation}
E^2 \approx p^2c^2 \left[1 + \sum_{n=1}^{\infty} S_n \left(\frac{E}{E_{\mathrm{QG},n}}\right)^n \right],
\label{eq:mdr}
\end{equation}
where $E$ and $p$ represent the photon energy and momentum, respectively, $E_{\mathrm{QG},n}$ denotes the quantum gravity energy scale, and $S_n = \pm 1$ encodes the sign of the violation. $S_n = +1$ would result in a photon group velocity greater than $c$ (the so-called \textit{superluminal} behaviour), and the opposite is valid for $S_n = -1$ (\textit{subluminal} behaviour). Different orders in the series are treated independently. The sensitivity of current measurements allows us to probe only the first two orders. 

The first-order modification ($n=1$) follows from theoretical models with CPT-odd operators. These allow vacuum birefringence, which was strongly constrained independently (see, e.g.,~\cite{Toma:2012xa, Kostelecky:2013rv, Gotz:2014vza}). Vacuum birefringence does not appear in the $n=2$ class of models. Moreover, superluminal behaviour allows photons to split to three gamma rays or to decay to electron-positron pairs, which was also strongly constrained~\cite{HAWC:2019gui, LHAASO:2021opi}. 

In this case, we restricted ourselves to second-order correction in the subluminal scenario ($n = 2$, $S_2 = -1$). 
This choice is motivated by several reasons: second-order ($n=2$) operators dominate in effective field theories, preserving CPT and gauge invariance. Furthermore, they yield both time-of-flight effects and modifications to reaction thresholds and cross sections. Additionally, the second-order modification is observationally less constrained than the first-order one. 
The specific effects that we have considered were the time-of-flight and modified absorption of gamma rays on extragalactic background light (EBL). 

The LIV-induced delay in the arrival of gamma rays from a source at redshift $z_s$, compared to the special relativistic case is given by 
\begin{equation}\label{eq:timedelay}
    \Delta t_2 = -\frac{3}{2}S_2 \frac{E^2}{E_{\mathrm{QG},2}^2} \kappa_2(z_s),
\end{equation}
where $\kappa_2(z_s)$ represents the distance contribution. We used the standard measure, first introduced in~\cite{DistancePiran}:
\begin{equation}\label{eq:kappa}
    \kappa_2(z_s) = \frac{1}{H_0} \int_0^{z_s} \frac{(1 + z)^2}{\sqrt{\Omega_\Lambda + \Omega_m (1 + z)^3}} \, dz.
\end{equation}
For the standard cosmological parameters of the $\Lambda$CDM cosmology model, we use $H_0 = 70\,\mathrm{km\,Mpc}^{-1}\,\mathrm{s}^{-1}$, $\Omega_\Lambda = 0.7$ and $\Omega_m = 0.3$.

The probability of gamma rays avoiding scattering on EBL and reaching Earth is given by
\begin{equation}\label{eq:probability}
    P(E, z_s) = \exp[-\tau(E, z_s)],
\end{equation}
where the opacity of the universe to gamma rays is given as
\begin{equation}\label{eq:opacity}
    \tau(E,z_s) = \int_{0}^{z_{\mathrm{s}}}dz\,\frac{dl}{dz}\int_{-1}^{1} d\cos\theta \left(\frac{1-\cos\theta}{2}\right) \int_{\omega_\text{th}(E,\theta)}^\infty d\omega \; n(\omega,z) \,\sigma(E(1+z),\omega,\theta). 
\end{equation}
Here, $n$ is the EBL spectral density of the background radiation, and $\omega_\text{th}$ the reaction threshold of the soft photon. LIV affects both $\omega_\text{th}$ and the cross section $\sigma(E(1+z),\omega,\theta)$. We stress that the expressions for cross-section usually used in the literature are approximations, and not particularly good ones. Therefore, we followed the recommendation from~\cite{PhysRevD.110.063035}, which offers the most complete calculation of the cross section for pair production in any LIV scenario to date.

\section{Data set and analysis}
\label{sec:analysis} 
The analysis method development and testing was performed on simulated data sets. These were based on the 2014 flare from Mrk~501 observed by the High Energy Stereoscopic System (H.E.S.S.) Collaboration~\cite{HESS:2019rhe}. Based on this data set, LIV studies of both time-of-flight and modified universe transparency were performed. Although these two LIV effects were tested independently, using this dataset as a sandbox will allow us to compare our results to the results of independent LIV tests. Moreover, both the spectrum and the light curve have rather simple functional forms, which reduces the extent of possible systematic effects. 

According to \cite{HESS:2019rhe}, the data set consisted of $\sim2000$ gamma rays between 1.3 and 20\,TeV. The intrinsic spectrum was well fitted with a power law with a slope of $\alpha = 2.03 \pm 0.04_\mathrm{stat} \pm 0.2_\mathrm{sys}$. 
The light curve at energies below 3.25\,TeV was fitted with two Gaussians with parameters $(A_1, \mu_1(\mathrm{s}), \sigma_1(\mathrm{s})) = (80.5 \pm 6, 2361 \pm 185, 2153 \pm 301)$, $(A_2, \mu_2(\mathrm{s}), \sigma_2(\mathrm{s})) = (60.5 \pm 11, 6564 \pm 220, 676 \pm 283)$, where $A_i$, $\mu_i$, $\sigma_i$ are the prefactor, mean, and width of each Gaussian, respectively.

\subsection{Artificial neural networks architecture}
\label{sec:ANN}
We employed two different models of neural networks. 
A \textit{basic} ANN consisting of the input layer with eight nodes, three hidden layers of 1000 nodes each, and the output layer of nine nodes. Each node in the input layer is reserved for one of the parameters that describe the spectrum or the light curve as detected on Earth. Of the nine output parameters, eight are reserved for parameters that describe the spectrum or the light curve at emission. The additional parameter represents $E_\mathrm{QG,2}$. All input and output parameters are mapped to a $[0, 1]$ interval. We use a simple mean square error to construct a cost function 
\begin{equation}
    \mathcal{L} = \frac{1}{2mn} \sum_{i,j=1}^{m,n} \left(\hat{y}_i^{(j)} - y_i^{(j)}\right)^2,
\end{equation}
where $m$ is the numeber of parameters in the output layer, $n$ is the total number of training examples, and $\hat{y}_i^{(j)}$ and $y_i^{(j)}$ are the reconstructed and true (targeted) values of parameters, respectively. 
    
As an alternative, a transformer, originally introduced by \cite{vaswani2017attention}, was chosen for its intrinsic ability to model non-local dependencies across long sequences. In this case, the input layer consists of 4000 nodes, representing the arrival time and energy of each of 2000 gamma rays in the data set. The output layer contains only two nodes, representing the value and the standard deviation of $E_\mathrm{QG,2}$. \\
With $\hat{\mu}_i$ and $\hat{\sigma}_i$ denoting the reconstructed mean and standard deviation of the LIV parameter for the $i$-th training sample, respectively, and $y_i = \log_{10}(E_{\mathrm{QG,2}})$ being the true (targeted) value, the cost function is defined as:
\begin{equation}
    \mathcal{L} = \frac{1}{n} \sum_{i=1}^{n} \left[ (\hat{\mu}_i - y_i)^2 + \left((\hat{\mu}_i - y_i)^2 - \hat{\sigma}_i^2\right)^2 \right].
\end{equation}
The first term penalizes errors in the mean prediction, while the second term enforces consistency between the squared residual and the predicted variance, encouraging well-calibrated uncertainty estimates. The predicted uncertainty $\hat{\sigma}_i$ is constrained to be non-negative by applying an absolute value to the second output neuron. 

A very useful aspect of transformers is the attention mechanism. The attention-based summary gives us information about which parts of the input sequence contribute the most to the model’s predictions. It is computed as:
\begin{align}
    \text{Attn}(Q, K, V) = \text{softmax}\left( \frac{QK^\top}{\sqrt{d_{\text{model}}}} \right) V,
\end{align}
where $d_{\text{model}}$ is the embedding dimension (128 in our case), $Q$ is the learnable query vector, $K$ are the keys, and, $V$ are the values (see~\cite{vaswani2017attention} for a detailed explanation). 
Analysing the attributed attention can be used as a sanity check, as we will see in Sec.~\ref{sec:results}.

\subsection{Data sets generation}
\label{sec:MonteCarlo}
Both models are trained on simulated data sets based on the 2014 flare from Mrk~501. 
Training data sets are generated by simulating approximately 2000 gamma rays. Each simulated gamma ray is described by an ordered pair $(E_e, t_e, P_e)$ representing the energy at emission, the emission time, and the survival probability. The last quantity is a real number from $[0, 1]$ interval obtained randomly from a homogeneous distribution. 
Photons are propagated through space considering the cosmological redshift and assuming various $E_\mathrm{QG,2}$, including $E_\mathrm{QG,2}\rightarrow\infty$, which corresponds to preserved Lorentz symmetry. Therefore, the detected quantities $(E_d, t_d, P_d)$ are given as
\begin{align}
    E_d &= E_e / (1 + z_s),\\
    t_d &= t_e + \Delta t_2,
\end{align}
where $\Delta t_2$ is given in Eq.~(\ref{eq:timedelay}), and $P_d$ in Eq.~(\ref{eq:probability}). Only gamma rays with $P_e > P_d$ will be detected. The final detected data set contains exactly 2000 gamma rays, and the size of the emitted data set varies, depending on $P_d$, which, in turn, depends on $E_\mathrm{QG,2}$. 
We assumed no correlation between gamma-ray energy and emission times and considered an ideal detector. 

Training sets for the basic ANN are created by fitting the distributions of $E_e$, $t_e$, $E_d$, and $t_d$. The two parameters describing the distribution of $E_d$ and the six parameters describing the distribution of $t_d$ constitute the input vector of each data set. The output vectors are composed of two parameters describing the distribution of $E_e$, six parameters describing the distribution of $t_e$, and one parameter for $E_\mathrm{QG,2}$. 

The training set for the transformer architecture consists of 7500 sets of 2000 gamma rays, each consisting of two numbers $E_d$ and $t_d$. Additionally, each set of gamma rays has a value of $E_\mathrm{QG,2}$, which acts as a target to which the transformer optimizes regression. The transformer receives 4000 values as the input vector, and outputs two numbers which represent the estimated $E_\mathrm{QG,2}$ and its standard deviation.

\section{Results and discussion}
\label{sec:results}
\begin{figure}[h]
\centering
\includegraphics[width=0.5\textwidth,clip]{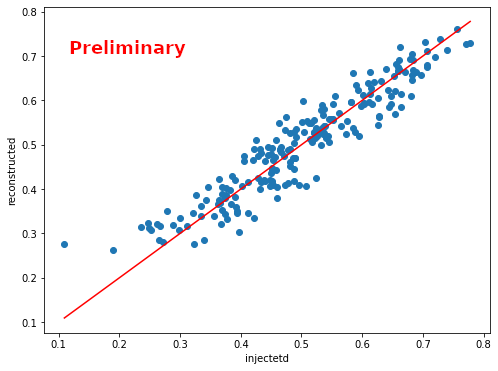}~~~
\includegraphics[width=0.5\textwidth,clip]{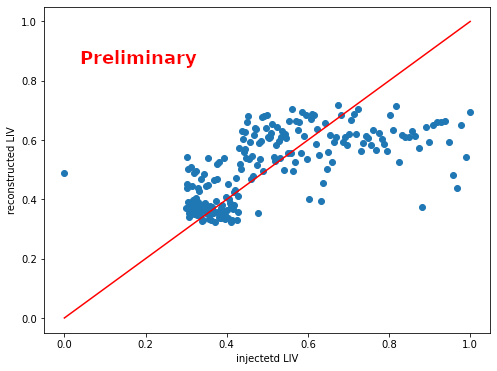}
\caption{Comparison of injected and reconstructed parameter values. \textit{Left:} Randomly chosen parameter from the light curve or spectral fit, with homogeneous distribution of values. \textit{Right:} LIV parameter. The red solid line shows the ideal reconstruction.}
\label{fig:basicANN}
\end{figure}

We evaluate the performance of ANNs on test samples, which were created in the same way and with the same characteristics as training samples. 
The results of the basic ANN are shown in Figure~\ref{fig:basicANN}. 
There is a strong correlation between the true and reconstructed values of the parameters that appear in both the emitted and detected samples and whose values are homogeneously distributed over the $[0, 1]$ interval. 
An example in the left plot of Figure~\ref{fig:basicANN}, has the Pearson correlation coefficient $(r, p) = (0.94, 10^{-98})$. 
On the other hand, the LIV parameter is not well reconstructed (right plot of Figure~\ref{fig:basicANN}). In this case, $(r, p) = (0.64, 10^{-25})$. An obvious issue is the fact that the $[0, 1]$ interval is not covered well. This part is somewhat tricky because the null hypothesis corresponds to an infinite value of $E_\mathrm{QG,2}$. For numerical reasons, it is represented by value 0, and there is a gap between the no-LIV case and the highest finite value of $E_\mathrm{QG,2}$. Therefore, special attention should be paid to seamlessly connecting this case with finite values of $E_\mathrm{QG}$. However, as we shall soon see, transformer architecture might offer an easier way out.

\begin{figure}[h]
    \centering
    \includegraphics[width=0.5\textwidth,clip]{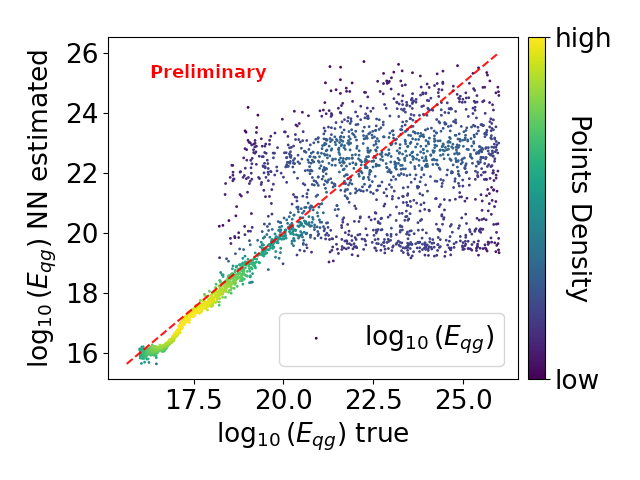}~~~
    \includegraphics[width=0.5\textwidth,clip]{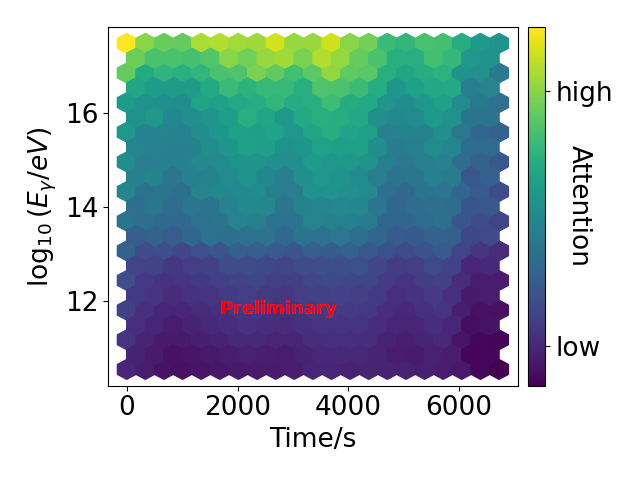}
    \caption{Performance of the transformer neural network. \textit{Left:} Comparison of injected and reconstructed values of $\log_{10}(E_{\text{QG}})$, coloured by density. The red dashed line indicates the ideal reconstruction. Values of $E_\mathrm{QG,2}$ are in eV.  
    \textit{Right:} Average attention weight across photons in the sequence as a function of detection time and photon energy.}\label{fig:TNN}
\end{figure}

The performance of the transformer is given in Figure~\ref{fig:TNN}. In the plot on the left, one can see a strong correlation between the injected and reconstructed values of $\log_{10}(E_{\text{QG,2}})$ up to $\sim 10^{10}$\,GeV. Beyond that value, the distribution of the reconstructed values is almost homogeneous, indicating a loss of prediction power beyond those scales. This does not come as a surprise because the constraints on $E_\mathrm{QG,2}$ set by the H.E.S.S. Collaboration, using the same data set, are $8.5\times 10^{10}$\,GeV based on time-of-flight, and $7.8\times 10^{11}$\,GeV based on modified gamma-ray absorption~\cite{HESS:2019rhe} (note that H.E.S.S. tested these two LIV effects as mutually independent). Therefore, our capabilities are limited by the data set. 
This result might be helpful with the problem of modelling Lorentz invariant case ($E_\mathrm{QG,2}\rightarrow\infty$) that we encountered in the basic ANN approach, that is, for energy scales beyond the sensitivity of the data set, effect of any values of $E_\mathrm{QG,2}$ is not recognisable from Lorentz invariance. 
However, without modelling a Lorentz invariant scenario, the question is how to estimate the upper limits on $E_\mathrm{QG,2}$. 

The attention map shows that more weight is given to the higher energy photons and photons in the $2 - 4$\,ks time interval. The former is perfectly expected, since the high-energy photons are more strongly affected by LIV than the low-energy photons. On the other hand, there is no obvious reason why more attention should be paid to a specific time interval, since LIV-induced time delay affects all photons equally, regardless of when they were emitted. One possible reason could be the fact that the Mrk~501 flare has two peaks, the first one centred around $2361 \pm 185$\,s containing more gamma rays (see Figure 1 in~\cite{HESS:2019rhe}), and consequently, dominating the analysis.

\section{Conclusion and outlook}
\label{sec:conclusion}

In this contribution, we presented a new approach to LIV studies. A fundamental novelty is the simultaneous testing of two coexisting LIV effects in the VHE gamma-ray data set. A technical innovation, on the other hand, is the application of ANN for this purpose. 
We investigated the application of two different deep learning architectures. A simple ANN successfully reconstructed some of the parameters that were present in both the input and output layers of the training samples. The parameter representing LIV was present only in the output layer, but in this case, the reconstruction power turned out to be substantially weaker. One clear issue is the proper modelling of the Lorentz invariant case. However, the results of the transformer showed a possible, easy solution to this problem. Sequence-to-regression transformer shows a promising reconstructive power up to a certain level of $E_\mathrm{QG,2}$, which is expected, knowing that the same data set was already used in previous, although simpler, LIV studies. Attention mapping is a neat feature, allowing for a sanity check of the transformer's performance. 

There is room for improvement in both approaches, particularly following more rigorous tests of both methods, which are our short-term plans for the continuation of this research. 
Our mid-term plans include investigating different approaches, e.g., using individual events or bins in the input layer, instead of fit parameters for the basic ANN approach. Furthermore, we will introduce realistic detectors with non-unit acceptance and finite energy resolution. That will be followed by an estimate of the sensitivity to systematic effects and a comparison of the sensitivity to other analysis methods, such as likelihood. 
In the long term, we intend to combine multiple observations, introduce additional LIV effects (e.g., extensive air shower development), introduce additional free parameters representing source intrinsic correlation, cosmology, background models, etc.

Testing several coexisting LIV effects in a single analysis is a requirement posed by the underlying theories. However, one should not expect it to increase the sensitivity of the analyses. If anything, combining different effects introduces additional degrees of freedom, which in turn increases systematic effects and possibly introduces degeneracies. Nevertheless, this approach is more realistic from the physical point of view, and the results thereof are expected to be more robust.

\acknowledgments
The authors acknowledge the networking support by the COST actions CA18108 (QG-MM, \url{https://qg-mm.unizar.es/}) and CA23130 (BridgeQG, \url{https://web.infn.it/BridgeQG/}), and the financial support from the University of Rijeka through the uniri-iskusni-prirod-23-24 project and the Croatian Science Foundation (HrZZ) Project IP-2022-10-4595.
We thank the Institute for Fundamental Physics of the Universe (IFPU, \url{https://www.ifpu.it/}) for hosting the workshop ``Astrophysical searches for quantum-gravity-induced time delays'' where ideas important to this work were developed.

\bibliographystyle{ieeetr}
\bibliography{ICRC2025_863}

\begin{thebibliography}{10}

\bibitem{AmelinoCamelia:1997gz}
G.~Amelino-Camelia, J.~R. Ellis, N.~E. Mavromatos, D.~V. Nanopoulos, and S.~Sarkar, ``{Tests of quantum gravity from observations of gamma-ray bursts},'' {\em Nature}, vol.~393, pp.~763--765, 1998.

\bibitem{LHAASO:2021gok}
Z.~Cao {\em et~al.}, ``{Ultrahigh-energy photons up to 1.4 petaelectronvolts from 12 $\gamma$-ray Galactic sources},'' {\em Nature}, vol.~594, no.~7861, pp.~33--36, 2021.

\bibitem{Addazi:2021xuf}
A.~Addazi {\em et~al.}, ``{Quantum gravity phenomenology at the dawn of the multi-messenger era{\textemdash}A review},'' {\em Prog. Part. Nucl. Phys.}, vol.~125, p.~103948, 2022.

\bibitem{PerezdelosHeros:2022izj}
C.~P\'erez de~los Heros and T.~Terzi\'c, ``{Cosmic Searches for Lorentz Invariance Violation},'' {\em Lect. Notes Phys.}, vol.~1017, pp.~241--291, 2023.

\bibitem{Terzic:2021rlx}
T.~Terzi\'c, D.~Kerszberg, and J.~Stri\v{s}kovi\'c, ``{Probing Quantum Gravity with Imaging Atmospheric Cherenkov Telescopes},'' {\em Universe}, vol.~7, no.~9, p.~345, 2021.

\bibitem{Toma:2012xa}
K.~Toma {\em et~al.}, ``{Strict Limit on CPT Violation from Polarization of Gamma-Ray Burst},'' {\em Phys. Rev. Lett.}, vol.~109, p.~241104, 2012.

\bibitem{Kostelecky:2013rv}
V.~A. Kosteleck\'y and M.~Mewes, ``{Constraints on relativity violations from gamma-ray bursts},'' {\em Phys. Rev. Lett.}, vol.~110, no.~20, p.~201601, 2013.

\bibitem{Gotz:2014vza}
D.~Gotz, P.~Laurent, S.~Antier, S.~Covino, P.~D'Avanzo, V.~D'Elia, and A.~Melandri, ``{GRB 140206A: the most distant polarized Gamma-Ray Burst},'' {\em Mon. Not. Roy. Astron. Soc.}, vol.~444, no.~3, pp.~2776--2782, 2014.

\bibitem{HAWC:2019gui}
A.~Albert {\em et~al.}, ``{Constraints on Lorentz Invariance Violation from HAWC Observations of Gamma Rays above 100 TeV},'' {\em Phys. Rev. Lett.}, vol.~124, no.~13, p.~131101, 2020.

\bibitem{LHAASO:2021opi}
Z.~Cao {\em et~al.}, ``{Exploring Lorentz Invariance Violation from Ultrahigh-Energy \ensuremath{\gamma} Rays Observed by LHAASO},'' {\em Phys. Rev. Lett.}, vol.~128, no.~5, p.~051102, 2022.

\bibitem{DistancePiran}
U.~{Jacob} and T.~{Piran}, ``{Lorentz-violation-induced arrival delays of cosmological particles},'' {\em J. Cosmology Astropart. Phys.}, vol.~1, p.~031, Jan. 2008.

\bibitem{PhysRevD.110.063035}
J.~M. Carmona, J.~L. Cort\'es, F.~Rescic, M.~A. Reyes, T.~Terzi\ifmmode~\acute{c}\else \'{c}\fi{}, and F.~I. Vrban, ``Approaches to photon absorption in a lorentz invariance violation scenario,'' {\em Phys. Rev. D}, vol.~110, p.~063035, Sep 2024.

\bibitem{HESS:2019rhe}
H.~Abdalla {\em et~al.}, ``{The 2014 TeV $\gamma$-Ray Flare of Mrk 501 Seen with H.E.S.S.: Temporal and Spectral Constraints on Lorentz Invariance Violation},'' {\em Astrophys. J.}, vol.~870, no.~2, p.~93, 2019.

\bibitem{vaswani2017attention}
A.~Vaswani, N.~Shazeer, N.~Parmar, J.~Uszkoreit, L.~Jones, A.~N. Gomez, L.~u. Kaiser, and I.~Polosukhin, ``Attention is all you need,'' in {\em Advances in Neural Information Processing Systems} (I.~Guyon, U.~V. Luxburg, S.~Bengio, H.~Wallach, R.~Fergus, S.~Vishwanathan, and R.~Garnett, eds.), vol.~30, Curran Associates, Inc., 2017.

\end{thebibliography}

\end{document}